\documentclass[12pt]{article}
\usepackage{epsfig}

\parskip=\medskipamount 
plus.3em minus.5em \abovedisplayshortskip=.5em plus.2em minus.4em
\renewcommand{\thesection}{\arabic{section}}

\catcode`@=11

\@addtoreset{equation}{section} \@addtoreset{equation}{subsection}
\def\theequation{\ifnum\value{section}=0 \arabic{equation}\ignorespaces
\else \ifnum\value{section}=-1 A.\arabic{equation}\ignorespaces
\else \ifnum\value{subsection}=0
\thesection.\arabic{equation}\ignorespaces \else
\thesection.\arabic{subsection}.\arabic{equation}\ignorespaces
                             \fi
                        \fi
                   \fi}

{\catcode`\'=\active \def'{{}^\bgroup\prim@s}}

\catcode`@=12

\newcommand{\bq}{\begin{equation}}
\newcommand{\be}{\begin{equation}}
\newcommand{\fq}{\end{equation}}
\newcommand{\ee}{\end{equation}}
\newcommand{\bqr}{\begin{eqnarray}}
\newcommand{\beqs}{\begin{eqnarray}}
\newcommand{\fqr}{\end{eqnarray}}
\newcommand{\eeqs}{\end{eqnarray}}

\newcommand{\rf}[1]{(\ref{#1})}

\def\bop#1{\setbox0=\hbox{$#1M$}\mkern1.5mu
    \vbox{\hrule height0pt depth.04\ht0
    \hbox{\vrule width.04\ht0 height.9\ht0 \kern.9\ht0
    \vrule width.04\ht0}\hrule height.04\ht0}\mkern1.5mu}

\begin{document}
\thispagestyle{empty}

\begin{flushright}
\begin{tabular}{l}
hep-th/0507074 \\
\end{tabular}
\end{flushright}

\vskip .6in
\begin{center}

{\bf A Note on Singularities and Polynomial Zeros}

\vskip .6in

{\bf Gordon Chalmers}
\\[5mm]

{e-mail: gordon@quartz.shango.com}

\vskip .5in minus .2in

{\bf Abstract}

\end{center}

The analysis of solutions to algebraic equations is further simplified.  
A couple of functions and their analytic continuation or root findings 
are required. 

\vfill\break

A simplified version of the solutions to algebraic equations is presented.  
In previous work, a generating function to sets of these equations is derived 
via the geodesic flow on potentially singular manifolds \cite{ChalmersOne}. 
The general solution to these geodesic equations dictates solutions to 
polynomial 
equations not only in the integer and rational fields, but also in irrational 
and transcendental ones.  The work \cite{ChalmersOne} is not reviewed here. 
There is formalism related to the work in \cite{ChalmersTwo} and 
\cite{ChalmersThree} 
both in number theory and an implementation in scalar field theory.  

The solution set to the elliptic equation in 

\bqr 
y^2=x^3+ax+b \ , 
\label{ellipticone} 
\fqr 
and in general the hyperelliptic equations, as well as systems of coupled 
polynomial equations, is of interest for many reasons.  The counting of these 
solutions can be done not only in the order set (the number) but of the 
number type. 

The solution set to the equation can be found from doubling the curve in 
\rf{ellipticone} with an anti-holomorphic counterpart, finding the metric 
on the space which is dependent only on $a$ and $b$, and generating the 
geodesic flows from a known solution such as $y^2=b$ and $x=0$ for $b$ 
equaling a square.  

The geodesic flow equations are represented by the two equations, 
\bqr 
m=f(\tau;m_0,n_0) \qquad n=g(\tau;m_0,n_0)  \ , 
\label{geodesicsolution} 
\fqr
together with their redundant conjugate equations.  These equations 
require the metric on the elliptic curve in order to analyze, and the 
singularites label all of the disallowed integers $m$ and $n$; the latter 
represent non-allowed solutions to the curve.  Systems of equations 
such as a hyperelliptic ones, or coupled equations in multi-variables 
can be analyzed in the same way.  

The differential representation in \rf{geodesicsolution} has a formal 
power series expansion, 

\bqr 
m= \sum a_p(m_0,n_0) \tau^p  \qquad n=\sum b_q(m_0,n_0) \tau^q \ .  
\label{series}  
\fqr 
This series has to be analyzed, given the metric data, for the existence 
of a $\tau$ given a trivial solution such as $m_0=n_0=0$.  One condition is 
that one pair of the roots, pair meaning a root from both equations in \rf{series}, 
has to be real, non-zero, and identical.  The coefficients $a_p(m_0,n_0)$ and 
$b_q(m_0,n_0)$ are derived from the metric data on the four-dimensional holomorphic 
and anti-holomorphic elliptic curves.  In general, however, the full functional 
dependence on $\tau$ in \rf{series} should be analyzed including a possible 
analytic continuation (e.g. hypergeometric functions for example).      

The group of integers are set into the polynomial equation 

\bqr 
Q(x,m)=\prod \bigl(x-\sum_{n=0} h(m,n)\bigr) \ .  
\label{disallowedint}
\fqr 
Another polynomial $P(x,m)$ can be defined which represents the complement 
set of integers, i.e. the allowed solution set to the elliptic equation, 

\bqr 
P(x,m)={\prod_{m=0} (x-m)\over Q(x,m)} \ .   
\label{allowedsolutions} 
\fqr 
The polynomial in \rf{allowedsolutions} generates all of the integer, or 
in general rational number solutions to the the curve, 

\bqr 
y^2=x^3+ax+b \ . 
\label{elliptictwo}   
\fqr 
The degree of the polynomial $P(x,m)$ in \rf{allowedsolutions} generates 
the order of the solution set to \rf{elliptictwo}.  

Basically the functions 

\bqr 
G^{(1)}_{m,n;m_0,n_0}= \sum a_p(m,n) \tau^p-m_0  
\fqr 
and 
\bqr 
G^{(2)}_{m,n;m_0,n_0}= \sum b_q(m,n) \tau^q -n_0 \ .  
\label{functions}  
\fqr 
or 
\bqr 
G_{m,n;m_0,n_0}=G^{(1)}+i G^{(2)} \ , 
\fqr 
are required for the number of zeros and singularities.  These functions are 
found from the metric on the four-dimensional space spanned by the elliptic 
curve in \rf{elliptictwo} and its anti-holomorphic counterpart.  These functions 
are generalized elliptic, or Appel functions, with a further generalization 
to non-elliptic coupled algebraic systems including hypergeometric ones.  
Standard differential or analysis techniques can find the zeros and 
infinities, as well as their analytic continuations.  The number of 
zeros of the functions, or of the complex one $G_{m,n;m_0,n_0}(\tau)$, are required to find the polynomial solutions to 
the elliptic curve in \rf{ellipticone} and group them into the function 
in \rf{allowedsolutions}.  The latter function seems clear to derive 
as well as its order.  

The metric on these spaces is direct to find through a tree-level D-term 
analysis of the quiver formulation \cite{ChalmersFour}.  These metrics are 
found number theoretically through the pinching of classical $\phi^3$ diagrams, 
through counting their number at zero momentum \cite{Unpublished}.  The 
geodesic flow solutions generating these functions are one-dimensional 
and are derivable \cite{Unpublished}.  General 
systems of polynomial equations are mapped to sets of functions in \rf{functions} depending on their number and dimensionality, i.e. variables in $x$ and $y$.  The analytic continuation, and their number of zeros and infinities, count the solutions 
in various number fields. 

\vfill\break

\end{document}